\documentclass[a4paper,11pt]{article}
\usepackage{adjustbox}
\usepackage{xcolor}
\begin{document}
	\title{Our Dark Matter Stopping in the Earth }
	\author{H.B. Nielsen\footnote{Speaker at the  Work Shop
			``What comes beyond the Standard Models'' in Bled.}, Niels Bohr Institut,\\
		and Colin D. Froggatt, Glasgow University}
	\date{``Bled''   , July , 2023}
	\maketitle
	\begin{abstract} We have worked for some time on a model for dark matter,
		in which dark matter consists of small bubbles of a new speculated type of vacuum,
		which are pumped up by some ordinary matter such as diamond, so as to
		resist the pressure of the domain wall separating the two vacua. Here we
		put forward thoughts on, how such macroscopic pearls would have their surrounding dust cleaned off
		passing through the atmosphere and the Earth, and what their distribution would be as a function of the depth of their stopping point and the
		distribution of the radiation emitted from them. In our model we assume that
		they radiate 3.5 keV electrons and photons, after having
		been excited during their passage into the Earth. The purpose of such an estimation of the radiation distribution is to explain the truly mysterious fact that, among all the underground experiments seeking dark matter colliding with the Earth material, only the DAMA-LIBRA experiment has
		seen any evidence of dark matter. This is  an experiment based on 
		{\em solid NaI} scintillators and is
		rather deep at 1400 m. It is our point that we can arrange the main
		radiation to appear in the relatively deep DAMA-LIBRA site, and explain that
		the dark matter pearls cannot stop in a fluid, such as xenon in the xenon
		based experiments. 
	\end{abstract}
	\section{Introduction}
	
	It is still a great {\em mystery} of what the {\em dark matter}, of which one mostly has
	seen the effect of its gravitation, {\em consists}. An exceptionally great
	mystery in
	this connection is that among the experiments looking for dark matter
	hitting the Earth and being detected deep underground - to avoid the
	cosmic radiation background - there is only one experiment, DAMA-LIBRA \cite{DAMA2}
	which has seen any evidence for dark matter. The experiments based on the fluid
	scintillator, fluid xenon, typically claim direct disagreement with
	DAMA-LIBRA, by obtaining so low upper limits on the cross section
	for the dark matter - using a WIMP model - that the observations by DAMA could
	not avoid having been seen in e.g. LUX. Our model has dark matter, that is
	not as weakly interacting as the usual WIMP model assumes. Rather our
	dark matter pearls consist of bubbles of a new (speculated) type of vacuum, containing ordinary matter and
	compressed to an outrageously high density. So, although still
	being per kg much less interacting than ordinary matter, our dark matter is
	much more strongly interacting per kg than the WIMPs usually considered.
	Thus our pearls of dark matter should not be called WIMPs but rather only
	IMPs (Interacting Massive Particles). The essential point for the present 
	paper is, what happens to our model dark matter when hitting the Earth. It  is
	not so much that they consist of a new type of vacuum etc. but rather
	that they are macroscopic objects causing a much bigger interaction that
	matters. However they
	still must be so massive compared to their cross section, that they
	do not just function like ordinary matter. But that does not prevent 
	them getting stopped in the Earth, although with an appreciably longer stopping length than ordinary matter.
	
	It is rather easy in our model to adjust parameters, so as to obtain
	whatever stopping length one might want for our dark matter pearls. At least we can easily believe that
	we could fit them to have a stopping length of the order of the depth of the
	DAMA-LIBRA experiment. Then if they were arranged to emit most of their
	excitation energy where they get stopped, they
	would be appreciably more visible to experiments in the depth of DAMA-LIBRA than in other depths. 
	
	If really the stopping of dark matter is needed for their easy observation, then the experiments
	with a fluid scintillator would be severely disfavoured because a dark matter
	pearl cannot really fully stop in a fluid. A little piece of fluid around the
	pearl will at least by gravity follow the pearl as it falls down, and it would
	spend much less time in a liquid xenon experiment than in a NaI(Tl) one. Even
	other NaI(Tl) experiments, if having a lower depth under the Earth surface
	than DAMA, might see only a little of the dark matter, because it passes through such experiments too fast
	and too little of the radiation from its excitation would be detected at such higher up experiments.
	
	So our model of dark matter consists of small but still macroscopic pearls, in the sense of each consisting
	of many atoms, then made effectively ``darker'' by having these atoms
	concentrated by the domain wall between the new vacuum and the ordinary vacuum, which we assume to be degenerate (i.e. with the same energy density). We shall first briefly mention the evidence from dwarf galaxies for the interaction of dark matter with itself.

	\section{ Our dark matter, stopping in the Earth, etc.}
	
	
	
	The two most crucial properties of our model      \cite{Dark1,Dark2,Tunguska,supernova,Corfu2017,Corfu2019,theline,Bled20,Bled21,
		extension,Bled22,Corfu2022,Corfu2022A} are as follows.
	
	\begin{itemize}
		
		\item Our dark matter is not so dark as WIMPs. Our dark matter pearls interact so much that they get stopped in the Earth.
		
		We actually speculate that our dark matter pearls
		get essentially stopped in the Earth, down from their $\sim 300km/s$ galactic velocity, at a 
		depth around that of DAMA-LIBRA. 
		Velocity dependent fits of their ``inverse darkness'' $\frac{\sigma}{M}$   by Correa \cite{CAC} from studying
		dwarf Galaxies is seen in Figure \ref{f1}, which was motivated by the failure of the model
		that dark matter has only gravitational interactions as seen in
		Figure \ref{gravitationonly}.
		\begin{figure}  
			\includegraphics[scale=0.9]{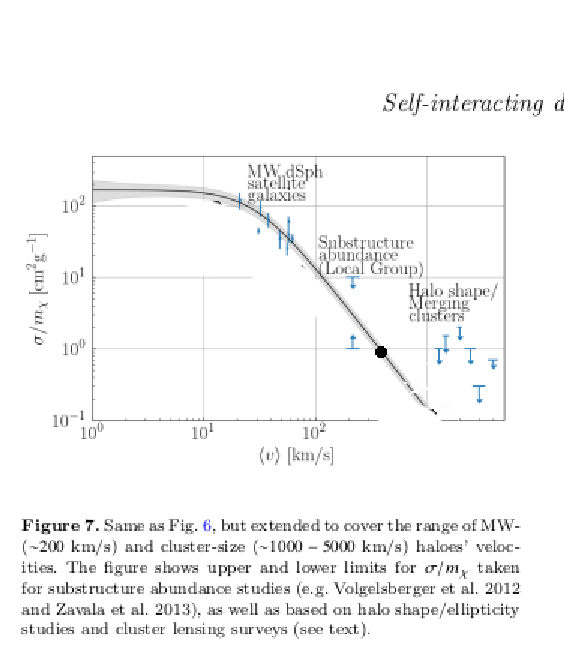}
			\caption{\label{f1} Self-interaction cross section over mass of the particles
				(pearls in our model) obtained by fitting velocities of (of course) the
				ordinary matter in dwarf galaxies around our Milky Way.}
		\end{figure}
		
		From Figure 2 we obtain the ``inverse darkness'' values:
		\begin{eqnarray}
			\frac{\sigma}{M}&\approx& 150 cm^2/g = 15 m^2/kg\, \hbox{(low velocity)}
			\label{id15}\\
			\frac{\sigma}{M}&\approx& 1.5 cm^2/g = 0.15 m^2/kg\, \hbox{(200km/s)}
			\label{id0k15}
		\end{eqnarray}
		But here we must admit that with these ``inverse darknesses'' the pearls
		would not go so deep as we would like to make them stop in the depth
		of the DAMA-LIBRA experiment. So we have to help our model a little bit
		by declaring that the pearls have caught up dust around them so as to get
		less dark, but that in the atmosphere or the upper shielding of the Earth 
		this dust is washed off. Washing off the dust makes the cross section
		smaller presumably without losing much weight, so that after such a
		cleaning the inverse darkness is decreased and thus the stopping length is
		increased.
		
		\item The dark matter particles can be excited to radiate 3.5 keV X rays or presumably also electrons of the same energy per particle.
	\end{itemize}
	
	
	\begin{figure}
		\includegraphics[scale=0.9]
		{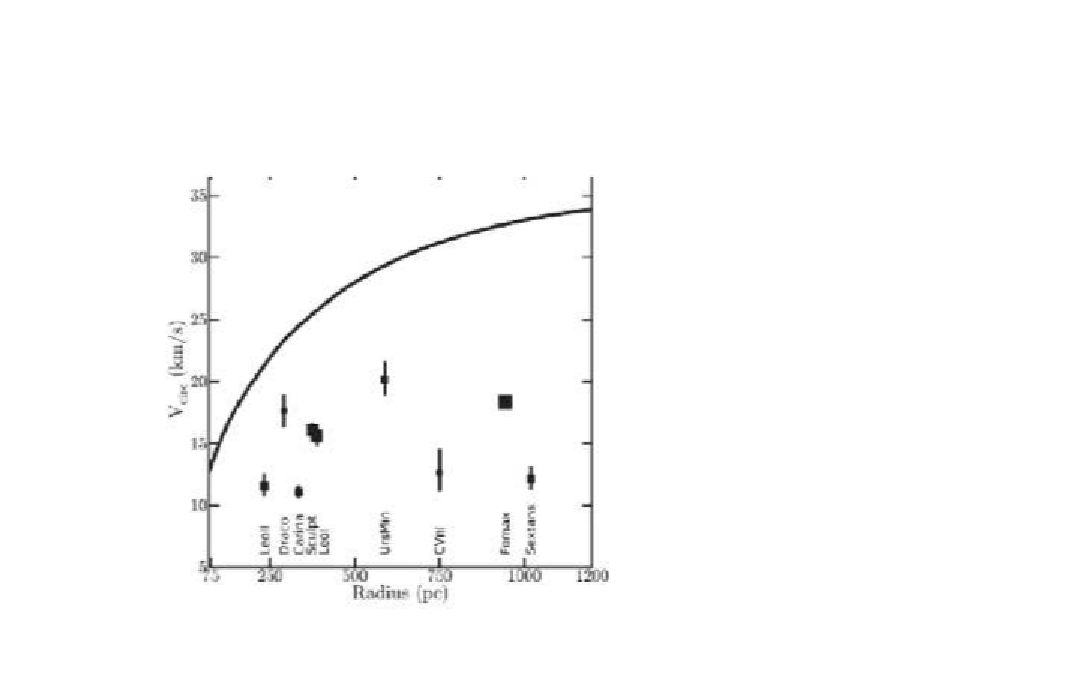}
		\caption{This figure should illustrate, that when Correa
			simulated the dark matter distribution, under the
			assumption of a purely gravity interaction, and regained a prediction
			for the star velocities it was {\em not} successful.
			So there is a phenomenological call
			for e.g. interaction of the dark matter with itself.}
		\label{gravitationonly}	
	\end{figure}

	\section{ A hard problem: How can ANAIS disagree with DAMA-LIBRA?}
	
	The two underground experiments, ANAIS \cite{ANAIS} and DAMA-LIBRA \cite{DAMA2}, and also a third one
	COSINE-100 \cite{cos100b}, which has not yet reached sufficient accuracy to properly disagree with
	either of the first two, looking for dark matter are
	{\em very similar}:
	\begin{itemize}
		\item Both use NaI(Tl= Thallium),
		\item Both look for seasonal variation,
		\item Depth is 1400 m $\sim 4200\, m.w.e.$ for DAMA-LIBRA, while ANAIS is
		at 2450m.w.e.  (COSINE-100 has a similar depth to ANAIS) 
	\end{itemize}
	
	But ANAIS ``sees'' no dark matter yet and claims to  deviate by
	3 $\sigma$ from  DAMA-LIBRA, which sees 0.0103 cpd/kg/keV.
	Our idea to resolve this problem is that:
	
	{\em Dark matter passes quickly by ANAIS, but slows down or stops at DAMA-LIBRA,}
		see Figure \ref{idea}. If the DAMA-LIBRA signal is due to radiating dark matter pearls then this could solve the problem.
	

	
	The COSINE-100 group searched for an annual modulation amplitude in their data taken over 3 years in the 1-6 keV region with the phase fixed to the DAMA value of 152.5 days. They subtracted their calculated time dependent background and obtained a positive amplitude of $0.0067 \pm 0.0042$ cpd/kg/keV, which is to be compared with the DAMA-LIBRA amplitude of 0.0103 cpd/kg/keV. The large error on the COSINE-100 amplitude means that it is consistent with both the DAMA-LIBRA result and no seasonal variation at all. 
 But then
  they analysed their own COSINE-100 data in the ``same way as
  DAMA'' \cite{cos100} by subtracting a constant background taken to be the average rate over one year, and generated what they consider to be a spurious seasonal variation. In fact they found a modulation amplitude of the opposite phase $ -0.0441 \pm 0.0057$ cpd/kg/keV (the wrong season has over abundance). Interestingly the WIMP model could not obtain such a result, but it is possible with our dark matter pearls.

  In our model an experiment at some given depth, would have just those
  dark matter particles stop in the instrument, which have just the right
  velocity.

  

 \subsection{How slowdown can help and a huge day/night effect}
\begin{figure} 
	\includegraphics[scale=0.7]{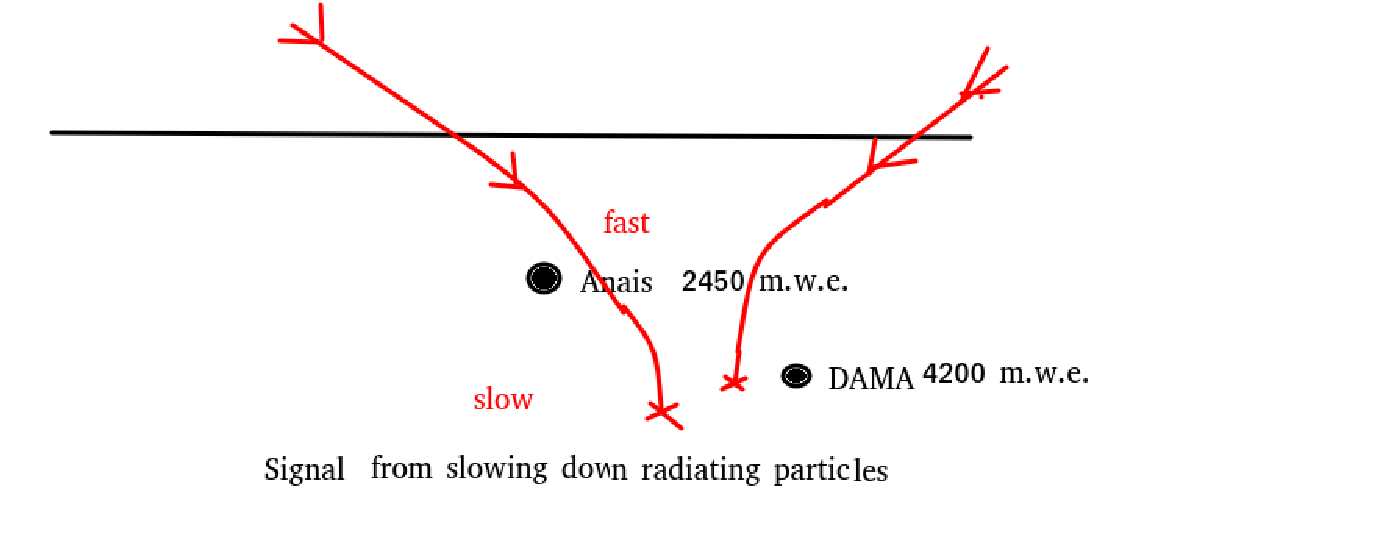}
	\caption{\label{idea} Under the assumption, that dark matter after
		excitation under the stopping radiates say electrons, and that it is these
		electrons, that have the best chance to be observed, the fast moving dark
		matter pearls in the region closer to the surface of the Earth leave
		only weak signals, while a much stronger signal will be left in the
		counters of DAMA-LIBRA in a depth, where the dark matter pearls have
		slowed down.(m. w. e. = meter water equivalent). Notice that we have
		drawn the trajectories as a bit curved, especially where the dark matter pearls have almost stopped, so much as to move slower. Then namely the effect
		of the Earth's gravity becomes relatively larger.}
\end{figure}

In Figure \ref{idea} we illustrate how one must think about
  dark matter particles coming in and at first having high speed but then slowing
  down due to the interaction with the earth or stone in the shielding.
  When the speed gets low even the Earth's gravity can make so much extra
  acceleration that the trajectories of the dark matter pearls become
  curved.

  But now we have to call attention to the fact that our model would - if we do not as we shall do in a moment find a way of avoiding it - predict a huge variation
  of the counting rate between day and night. The point is that the
  solar system moves with about 200 km/s relative to the dark matter center
  of mass system and, although the spread in the dark matter velocity in the
  radial direction to and from the center of the galaxy is presumably large,
  the spread in the direction around the galaxy in which the solar system moves
  will be much smaller. It might be as small as say $\pm 90$ km/s in which
  case the average of the dark matter velocity seen from the Earth or the Sun
  would be say 2 times as large as the spread. On the other hand with 
  stopping dark matter, as in our model, the dark matter can only come into
  the Earth and to an underground experiment from the side of the Earth on which the
  experiment is located. It can only come from the direction conceived as from
  above as seen from the experiment. This means that there will be a huge
  difference in the rate of visible impacts depending on whether the experiment
  in on the forward or the backward side of the Earth relative to the motion
  of the Earth seen from the dark matter average  rest system.
  
  When the experiment in question is on the front side of the Earth in its
  motion relative to the dark matter rest system, most of the dark matter
  can be overtaken by the Earth and fall into the experiment from the sky side,
  and one should get a very large rate of impact in this situation. Twelve
  hours later the experiment will be on the back side and now only very few
  dark matter particles have such a high individual velocity compared to the
  average velocity of the dark matter relative to the Earth, that they can run in
  the opposite direction to the majority of the dark matter. So in this case
  only relatively few dark matter particles can be observed.

\begin{figure}
	\includegraphics[scale=0.7]{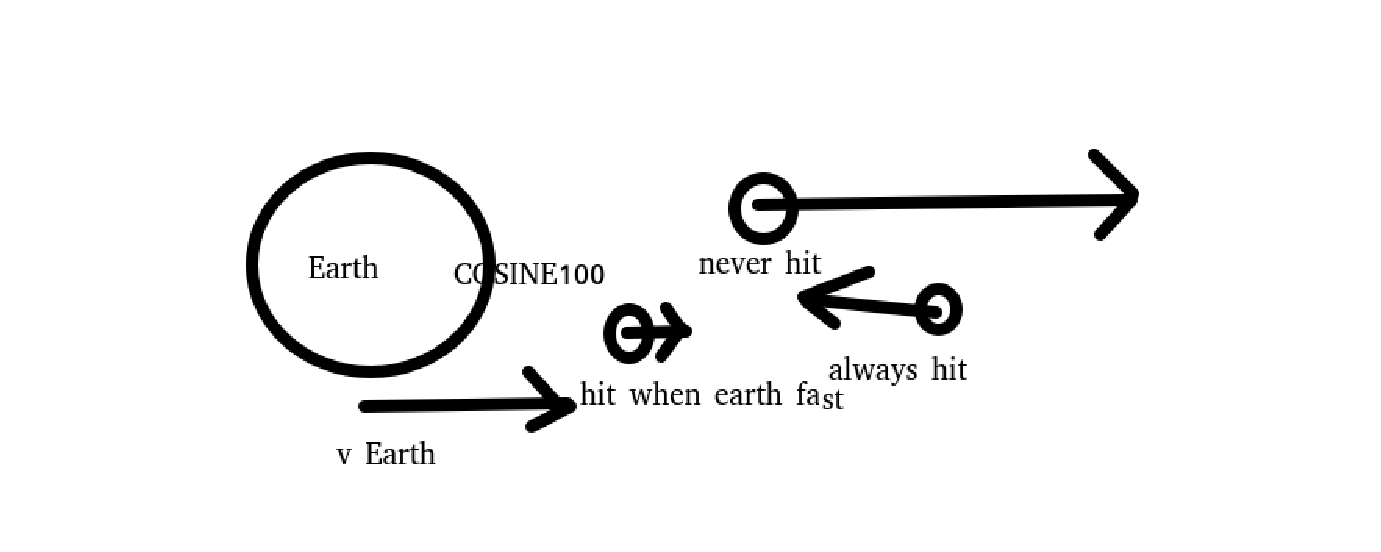}
	\caption{\label{f4} If the stopping length is so short as we speculate in
		our model, it is only dark matter that penetrates into the Earth from
		the side of the experiment, e.g. COSINE-100, that has any chance of being
		detected, while in WIMP-models dark matter particles coming through the
		earth-interior before interacting is not at all excluded. This figure
		concentrates on three examples of the velocity of particles having
		managed to come in on the front side of the Earth say, on which side the
		experiments lies. Whether dark matter
		particles now hit the earth, and how fast, depends on the relative velocity.}
\end{figure}

   For an underground experiment only particles running down towards its site can hit it, except that the Earth can overtake some slow ones. See Figure \ref{f4}. 

\begin{figure}
  	\includegraphics[scale=0.7]{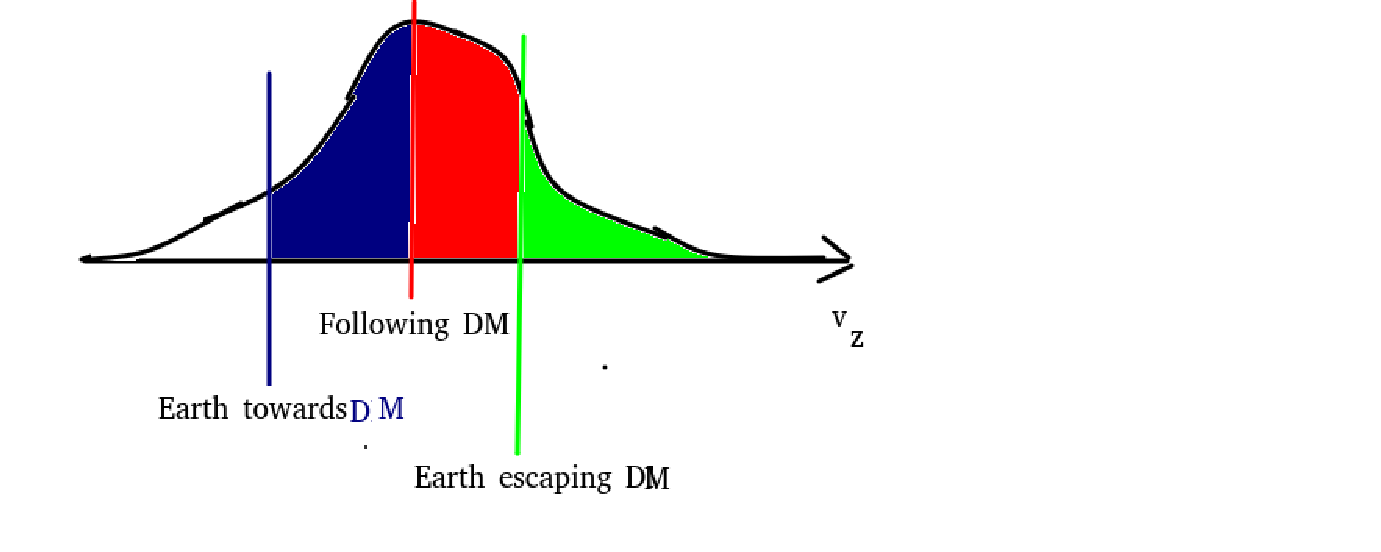}
  	\caption{\label{vd3} On this figure the Gaussian distribution is the
  		distribution of the dark matter particle velocities in the rest frame
  		for the average velocity of the dark matter, the center of mass. Let us
  		imagine the Earth at first coming in from the right side of the figure
  		with the site of the experiment, COSINE-100 say, on the place having
  		the left direction in zenith. Then, if the velocity towards the dark matter
  		of the Earth - moving to the left - is very high the dark matter in the
  		velocity bands denoted on the figure as dark blue, red and green, i.e.
  		all the ranges except the white one, would hit the Earth on the side of
  		this
  		experiment. But now if the Earth moved slower compared to the dark matter,
  		say it followed the center of mass for the dark matter, then only the
  		red and the green  amounts of dark matter would hit the Earth. And
  		if the Earth ``moved away'' in the sense that the experiment was on the
  		backside of the Earth compared to its velocity in the dark matter center
  		of mass frame, then say only the green part of the dark matter would hit
  		the Earth in the region of the experiment. In this case, where the experiment is
  		on the backside relative to the motion only the very fastest part,
  		the green band say, would be able to overtake the Earth and
  		hit the experiment.}
  \end{figure}

  In Figure \ref{vd3} we see the situation of the Earth running relative to the
  center of mass for dark matter. With the experiment in front of a huge
  part of the dark matter, all except the white tip would hit the Earth,
  while in the case of the experiment being on the backside only a small
  tip, the green one, will hit the Earth.
  
\begin{figure}  
	\includegraphics[scale=0.7]{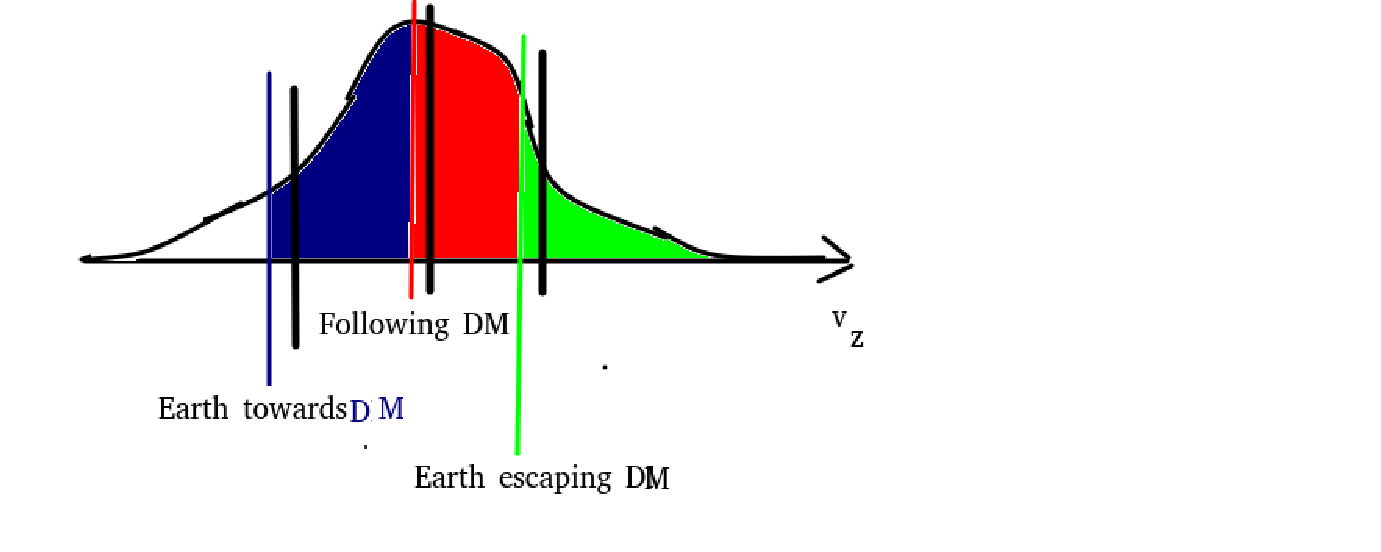}
	\caption{\label{vd3wbl} Like in figure \ref{vd3} we consider here a
		Gaussian dark matter velocity distribution, and look for how with different
		velocities of the Earth (with the experiment on the left side) moving in the
		dark matter sees more dark matter in the experiment the faster this
		Earth moves forward. 
		E.g. if the Earth moves with the experiment on the left side to the left,
		only the white
		part moves too fast for the Earth to catch up. But now we are interested
		in how small variations in the Earth's velocity, exemplified by the
		vertical lines drawn to the right of the color separation places on the
		figure. They symbolize a little bit slower Earth. The interesting point
		is that the effect of such a small velocity lowering is not the
		same independent of how fast the Earth moves already compared to the
		dark matter average. Even relatively the extra velocity gives different
		extra contributions, different even taken relative to the original
		one.}
\end{figure}

  In Figure \ref{vd3wbl} we have added some small displacement lines to
  illustrate how the borders between the amounts of dark matter hitting and not
  hitting the Earth is shifted by the relatively small velocity changes
  due to the season. When the experiment is in front in the motion towards the
  dark matter center of mass motion the hitting rate increases in the summer
  when the relative velocity is larger. However, when we are in the time
  of day when the experiment is on the backside the larger summer relative
  velocity will mean that the little tip of dark matter hitting the Earth gets even
  smaller. So in the time of day-night in which the dark matter mainly
  comes in from the side of the Earth opposite to that of the experiment,
  the seasonal effect is actually opposite to that when the dark matter comes in on the same side as the experiment, i.e
  there are most counts in the winter and fewer in the summer.

  This is definitely a very interesting prediction of ours, if the data on the
  detection of the dark matter is sufficiently detailed that one can distinguish
  day and night so to say. But now in detail our own model has it that
  the dark matter pearls are counted by means of the radiation of say
  electrons which are sent out delayed compared to the time at which the
  pearls were excited. If the decay lifetime is several days, then the
  very big day night oscillation gets washed out in our model and would not be
  seen. Indeed if the day night effect was as strong as we predict
  - percentwise stronger than the seasonal one - at first, i.e. if we did not
  have the washout by the pearls being excited and decaying with a several day
  lifetime scale, then it should have been easy for DAMA-LIBRA to have observed
  the day-night variation!
  (So it means that a model of our type, with only a limited penetration into the Earth, would not work unless we also have a washout due to
  the decay lifetime being of order of at least days.)

   When our dark matter pearls slow down, they either stop completely or continue falling slowly under gravity.
  \begin{itemize}
  \item If they stop completely, we should find a lot of stationary dark matter to be
    dug out as heavy pearls (much like if it were gold dust and one should wash
    it out).
    \item If they sink slowly, they of course go deeper inside the Earth. 
    \end{itemize}

  \section{Crude estimate of stopping distance and mass of a dark matter pearl}
\begin{figure}
      \includegraphics[scale=0.75]{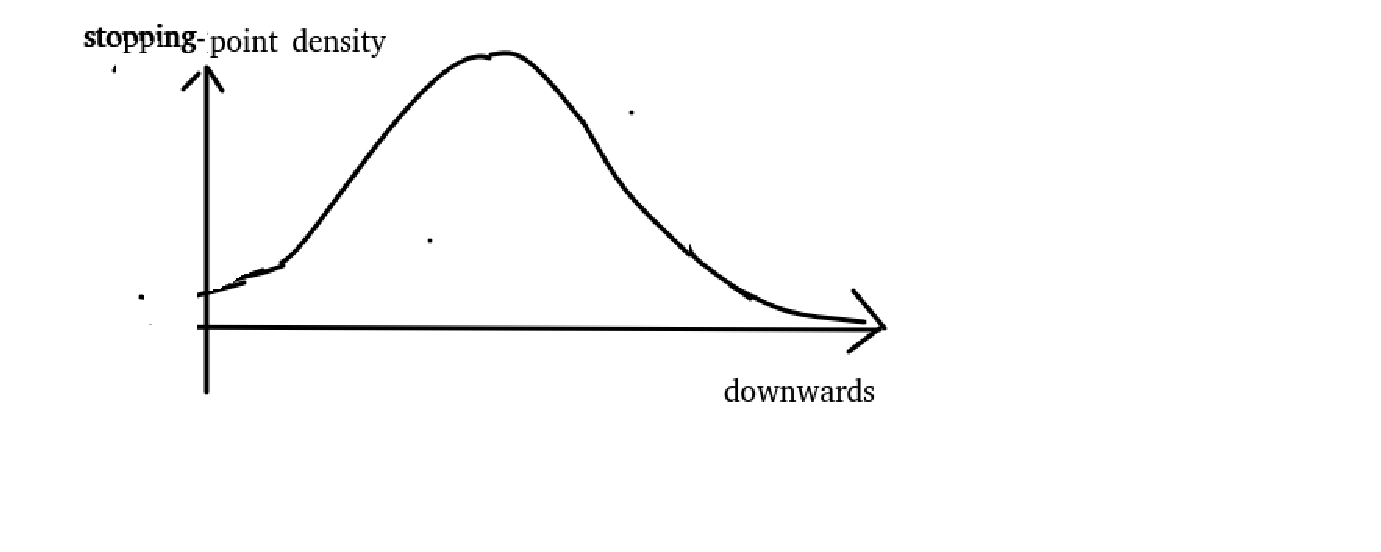}
      \caption{\label{fspd1} If one has a definite formula for the drag
        force stopping the pearls as a function of the velocity, then this
        formula will give the effective stopping depth - or rather the 
        penetration distance -
        as a function of the initial impact velocity (we consider only the vertical
        direction as an approximation). And then there will be simply a
        transformation from the impact velocity in the vertical direction to the
        depth at the point of stopping (only ``effectively'', if the particle really
        continues with a
        much lower fixed velocity, which is then neglected here).
        Here we have drawn symbolically the image under this transformation
        of the initial vertical velocity distribution. It still bears some
        similarity to the supposed Gaussian velocity distribution.}
  \end{figure}

  For simplicity we shall only consider the motion in one direction, down, and ignore the rest.
  The depth, into which a dark matter particle will penetrate before
  (effectively) stopping, will be a smooth function of its velocity relative
  to the Earth. So ``topologically'' the {\em distribution of stopping-depths}
  will reflect the {\em initial velocity spectrum} in the downward direction, as  
%
  illustrated in Figure \ref{fspd1}.
  
\begin{figure}   
  \includegraphics[scale=0.75]{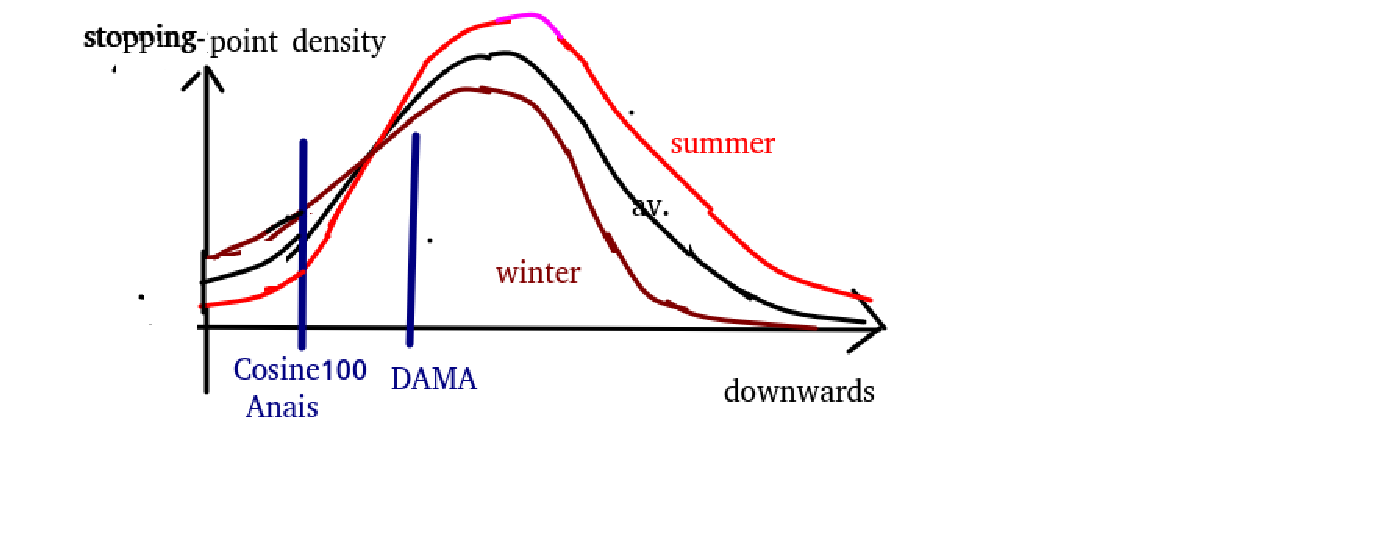}
  \caption{\label{fspd4ii} This is again the 
    vertical density distribution transformed into the depth spectrum (ignoring the horizontal components of the velocity), but now we have shown three variants
    corresponding to the seasons in which the  velocity of the dark
    matter relative to the Earth is different. In summer this relative velocity is higher and thus the stopping spectrum is for summer on average a bit deeper.
    On the contrary the winter spectrum lies a bit less deep. 
    An average spectrum is also drawn,
    which we could consider to be that for autumn and spring.
    Rather than to be realistic, the drawing has been made with the purpose of
    illustrating that indeed there is the possibility of the relatively
    close to the surface lying experiments ANAIS and COSINE-100 
    getting {\em less counts in the summer than in winter} opposite to the
    DAMA-LIBRA-experiment, which is about twice as deep (1400 m) and
    should see most events in the summer.}
    \end{figure}

   We note that DAMA-LIBRA is about twice as deep down as ANAIS and COSINE-100,
    see Figure \ref{fspd4ii}. This figure illustrates that in our model ANAIS and COSINE-100 could, in principle, have annual modulation amplitudes with the opposite phase to DAMA-LIBRA, but of course it is not necessary. 

  
 In order for our dark matter pearls to stop just at DAMA-LIBRA crudely, we need a penetration depth of the order of magnitude:
     \begin{eqnarray}
       \hbox{Penetration depth } L &=& \frac{M}{\sigma}*30 /
       \rho_{stone}\approx 1 km\\
       \Rightarrow \frac{M}{\sigma}&\approx&10^5 kg/m^2\\
       \Rightarrow \frac{\sigma}{M} &\approx& 10^{-5}m^2/kg
     \end{eqnarray}
   The number $30$ is a crude estimate involving a logarithm and factors of order unity roughly - see equation (\ref{Lstop}).

   We may compare this value of  $\frac{\sigma}{M}$ for reaching just the DAMA wanted order of magnitude
   with the numbers given in equations (\ref{id15}, \ref{id0k15}). This means that from the low velocity value  (\ref{id15}) we need a decrease by a factor $1.5 *10^6$ and from the $200 km/s$ value  (\ref{id0k15}) a decrease by $1.5*10^4$. Thus we must take it that
   so much dust around the dark matter pearl is washed off during the impact,
   that the cross section goes down by a factor of the order of respectively
   a million and 10000. Taking it that the relevant velocity is of the order of
   200 km/s, the factor ten thousand is enough.

   \subsection{Size of the bubble}

   The need for assuming the dirt to be washed off the pearl means that the inverse
   darkness $\frac{\sigma}{M} = 10^{-5}m^2/kg$, identified with the one needed for the penetration depth just reaching DAMA-LIBRA, must be that
   of the supposedly hard bubble making up the main and most heavy part of
   the dark matter pearls. Now we used to think that we could estimate
   the density of this bubble filled with ordinary matter under high
   pressure by using a dimensional argument \cite{theline} - of course rather uncertain - to obtain the Fermi-momentum from the HOMO-LUMO gap $E_H$ identified with the
   $E_H =3.5\, keV$ X-ray photon energy line likely to be associated with the dark
   matter:
   \begin{eqnarray}
     E_H  &=& \sqrt{2}\left (\frac{\alpha}{c}\right )^{3/2}E_f.
   \end{eqnarray}
   Here $E_f$ is the Fermi energy in the compressed ordinary matter inside
   the new vacuum bubble, and $\alpha$ is the fine structure constant, which
   for our a bit special dimensional argument has been taken to be of dimension
   velocity, so that it is $\alpha/c$ that is the usual fine structure
   constant $1/137.036...$. We actually even calculated the $\sqrt{2}$, but
   that would clearly be  far outside the expected accuracy. From this
   dimensional argumentation we then got the density of the material inside the
   new vacuum bubble:
   \begin{eqnarray}
     \rho_B &=& 5*10^{11}kg/m^3.
     \end{eqnarray}
   Of course the radius of the bubble with the compressed ordinary matter is
   given as
   \begin{eqnarray}
     R &=& \sqrt[3]{\frac{3M}{\rho_B*4\pi}}= \sqrt[3]{M}*8*10^{-4}kg^{-1/3}m.
    \end{eqnarray}
   and cross section
   \begin{eqnarray}
     \sigma =  \pi R^2& =&
     \pi^{1/3}3^{2/3}4^{-2/3}M^{2/3}\rho_B^{-2/3}\\
     &=& 1.21*(M/\rho_b)^{2/3} =  1.9*10^{-8}kg^{-2/3}m^2*M^{2/3}.
     \end{eqnarray}
  So we obtain
  \begin{eqnarray}
     \frac{\sigma}{M} &=& 1.9*10^{-8}kg^{-2/3}m^2 (\sqrt[3]{M})^{-1}
     \end{eqnarray}

 To achieve the value of $\sigma/M=10^{-5}m^2/kg$ to just reach DAMA we
 need

 \begin{eqnarray}
   10^{-5}m^2/kg &=&1.9*10^{-8}kg^{-2/3}m^2 M^{-1/3}\\
   \hbox{giving } \sqrt[3]{M} &=& 1.9 *10^{-3}kg^{1/3}\\
   \hbox{i.e. } M&=& 7*10^{-9} kg.
   \end{eqnarray}

 In a moment we shall see below, by considering the energy of impact of the dark matter, that we would for the purpose of what DAMA sees
 have preferred the value $9.4 *10^{-17} kg$.
   
   
    The density of dark matter in the region of the solar system is
     \begin{eqnarray}
       D_{sun} &=& 0.3\, GeV/cm^3.
       \end{eqnarray}
       The rate of impact energy is then
       \begin{eqnarray}
       ``Rate'' &=& v*D_{sun}\\
       &=& 300 km/s * 0.3 *1.79*10^{-27}kg/(10^{-6}m^3)\\
       &=& 1.6*10^{-16}kg/m^2/s\\
       &=&1.4*10^{-11}kg/m^2/day
       \end{eqnarray}
    

 Now let us call the mass of the individual pearls $M$ and then we can
 compute the number density of pearls of dark matter:
 \begin{eqnarray}
   \hbox{number density falling}&=&1.4*10^{-11}kg/m^2/day /M
   \end{eqnarray}
   Spreading over a kilometer we obtain: 
   \begin{eqnarray}
    \hbox{number density }&=&
   1.4*10^{-14}kg/m^3/day /M
   \end{eqnarray}
   For an earth density $= 3000kg/m^3$ we then have:  
   \begin{eqnarray}
   \hbox{number per kg of earth}&=& 4.7
   *10^{-18}/M per\; day\\
   \hbox{DAMA saw } S_M &=& 0.01 cpd/kg/keV\\
   \hbox{Over say 5 keV:\quad } 5keV S_m &=& 0.05 cpd/kg
   \end{eqnarray}
   If there is only one count per particle,we get
   \begin{eqnarray}
    4.7 *10^{-18}kg/M cpd &=& 0.05 cpd\\
   \hbox{giving } M&=& 9.4 *10^{-17}kg\\
   &=& 5.3 *10^{10}GeV 
 \end{eqnarray}
 \subsection{Discussion of mass value}

 The agreement of the mass $M=9.4 *10^{-17} kg $ needed to get the
 DAMA number observations with just one electron emission from each
 dark matter pearl and the number crudely estimated from the
 dimensional argument and the wish for the appropriate stopping length
 $7*10^{-9}kg$ is far from perfect. But we shall have in mind that the mass
 came in via a third root, so that a miscalculation of say $\rho_B$,
 the density of the compressed ordinary matter by just one order of magnitude
 would lead to a factor 1000 in the mass. This density $\rho_B$ again
 depends on the third power of the Fermi energy $E_f$, so that indeed
 the mass we need to achieve the wanted inverse darkness $\frac{\sigma}{M}
 = 10^{-5}m^2/kg$ will go as the ninth power, if there is a mistake in our
 $E_f$ estimate. So our 8 orders of magnitude deviation in the mass $M$
 corresponds to $E_f$ only being about one order of magnitude wrong.

 But our model is tensioned in the direction that the density of the
 bubbles should go up  and we should preferably get several counts
 out of the same dark matter pearl. It should sit in the NaI(Tl) and radiate
 for say several days. Then we could tolerate somewhat heavier pearls and still get
 the required number of counts.

 So in reality we can consider our model to be successful, because the
 deviations were not more than about 8 orders of magnitude in the mass, corresponding to one order of magnitude in the Fermi energy.

\section{Model}


We now give a short review of our dark matter pearl model:

    \begin{itemize}
    \item Our ``new physics'': There are several different phases of the
      vacuum, but all with same energy density ({\em Multiple Point
        Criticality Principle = MPP} \cite{MPP1,MPP2,MPP3,MPP4,tophiggs,Corfu1995,Dvali}). But even this ``several vacuum phases'' hypothesis
      is not truly new physics, if we believe the speculation that the quark
      masses have been adjusted so that two phases of the QCD-theory
      can be degenerate \cite{Corfu2022}:
      One phase with chiral symmetry spontaneously broken, and another one
      where the breaking should rather be said to be due to the quark masses.

    \item Our dark matter particles are macroscopic objects consisting
      of bubbles of a second vacuum filled with some ordinary matter e.g.
      carbon.
    \item There is a skin/wall separating the two vacuum phases with a
      tension of the order of
      \begin{eqnarray}
      	S^{1/3} &\sim& 10 \, MeV.
      \end{eqnarray}
      
    \item The nucleons have a lower potential in the inside-the-pearl-vacuum
      than
      in the outside vacuum, wherein we live, by a difference
      \begin{eqnarray}
        \Delta V &\approx& 3 \, MeV.
      \end{eqnarray} 

  
  The value $\Delta V \approx 3 \, MeV$ 
    was fitted to the inside material having a gap - a homolumo gap -
      between the empty and the filled electron-states arranged to let the
      dark matter preferably emit X-rays with the observed energy of
      3.5 keV.
    \end{itemize}
    

      \begin{itemize}
      \item The dark matter pearls are so macroscopic and of such a size compared to
        mass that they get stopped in the Earth at a distance of the order of the depth of the DAMA-LIBRA experiment (1400 m stone).
      \item The pearls are excitable so as to radiate X-ray photons of just the energy needed to give the ``mysterious 3.5 keV line''. 
        \end{itemize}

      \subsection{The 3.5 keV line}
      
      The observation of a mysterious new X-ray emission line at 3.5 keV was reported \cite{Bulbul,Boyarsky} in 2014. It was detected in the Andromeda galaxy, the Perseus cluster and different combinations of other galaxy clusters. Later the line was detected in the Milky Way Center \cite{Boyarsky2}.
      This line has been suggested to originate from dark matter and our model has been adjusted to fit the observations. However this interpretation  and even the very existence of the line is controversial. There are indeed several 
      details in the observation of this 3.5 keV X-ray, which a priori does not look
      supportive for the hypothesis that this line really comes from dark
      matter: In fact Jeltema et al. \cite{Jeltema} have seen it in the Kepler Supernova
      Remnant where there is far too little dark matter for them to be
      able to see it, unless dark matter interacts not only with dark matter but also
      with ordinary matter.
      Also the distribution seen from the Center of the Milky Way
      does not a priori look so much like dark matter produced X-rays.
      Furthermore there are some problems with the details of fitting the
      3.5 keV radiation from the outskirts of the Perseus galaxy cluster.
      Assuming that
      the dark matter can be brought to radiate the X-rays by interaction
      not only with other dark matter particles, but also with ordinary
      matter, we believe we can improve the understanding of these
      mentioned three problems in fitting the 3.5 keV radiation
      observations with dark matter pearls.
      
      We should point out that the observations of the 3.5 keV line are only seen as very small deviations
      of the spectra observed from what is understood from the various ions in the sources.
    They are  on the borderline
    of being statistically significant (see Figure \ref{Perseus} for the Perseus Cluster and Figure \ref{Andromeda} for Andromeda taken from from Iakubovskyi \cite{Iakubovskyi}):
      
  \begin{figure}    
    \includegraphics[scale=0.7]{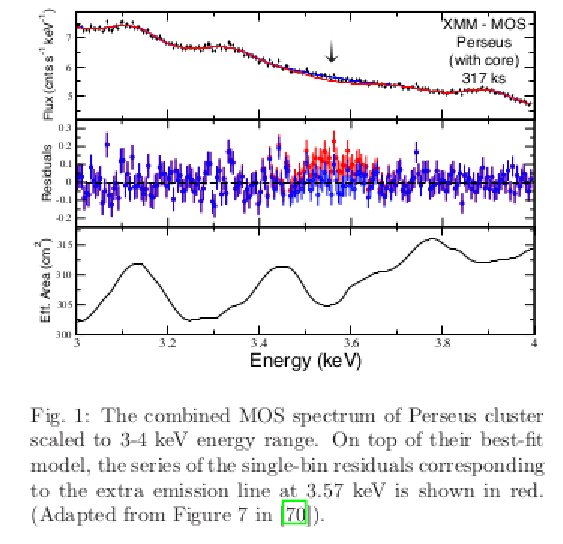}
    \caption{\label{Perseus} Spectrum of X-rays from the galaxy cluster
      Perseus. The curve is fitted to the expected X-ray spectrum from
      known ions. The little deviation at 3.5 keV could be from dark matter?}
    \end{figure}

  \begin{figure}
    \includegraphics[scale=0.8]{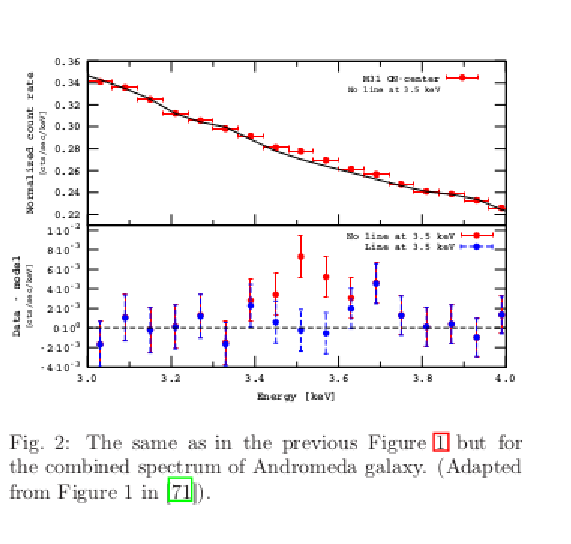}
    \caption{\label{Andromeda} X-ray spectrum from Andromeda.}
    \end{figure}

\begin{figure}
    \includegraphics[scale=0.3]{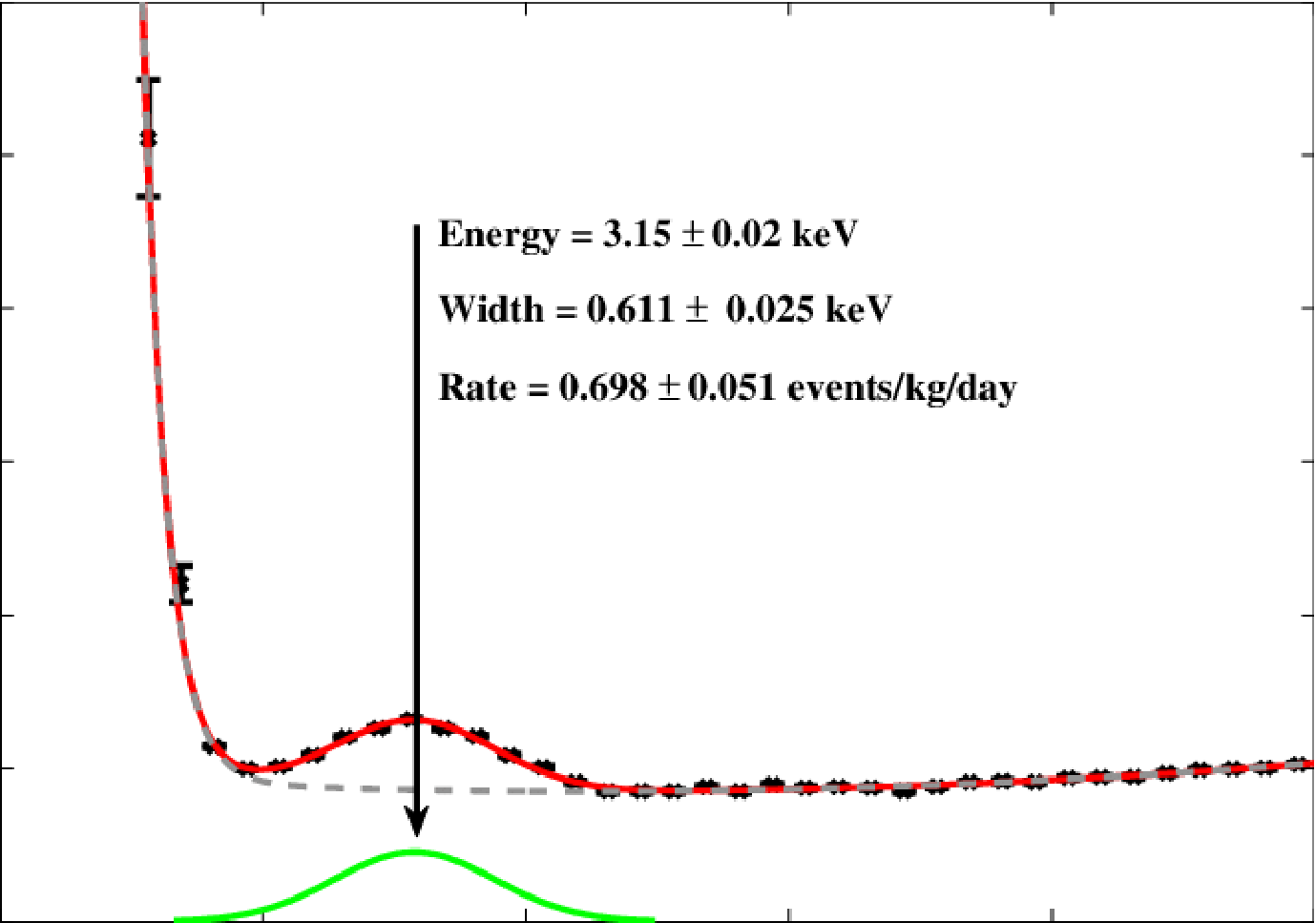}
    \caption{\label{DAMAspectrum} Here the seasonally varying part of the
      DAMA-LIBRA data is fitted to a Gaussian with the average energy 3.15 keV. This
      is remarkably close to the line observed from Perseus Cluster and
      Andromeda and other galaxy clusters etc.} 
    \end{figure}

  We identify the emission of some particle or another as seen in
  DAMA-LIBRA and potentially in COSINE-100 etc. with the emission from the
  mysterious 3.5 keV line. So it becomes of course crucial that the energy
  seen in the events of the seasonal varying type in DAMA-LIBRA is indeed
  just 3.5 keV. In Figure \ref{DAMAspectrum} we show a fit (red) to the  DAMA-LIBRA low-energy spectrum \cite{CDMSfit} consisting
  of a background model (grey/dashed) and a Gaussian distribution function
  (green). The parameters of the Gaussian are shown in the figure and the energy is given as 3.15 keV in remarkably good agreement!
  

\section{Distributions}

Really to get an idea of how the radiation from excited dark matter
pearls get distributed in the Earth, we must take into account that
the distribution of their incoming direction is presumably very smoothly distributed
over the sky.
We shall now make a couple of somewhat special
approximations and derive the distribution at the depth of the
point under the surface of the Earth at which one detects the radiation.

We shall make 4 ideal calculations based on the following assumptions:

\begin{itemize}
\item What is counted is the emissions from the dark matter particles of
  some radiation (electrons and/or X-rays, actually with energy 3.5 keV), which
  is assumed to go on with a constant rate for a long time (we may think of weeks)
  after the impact of the pearl. This means that in order to obtain
  a distribution with respect to depth, we need to calculate the length of time
  during which a dark matter particle is in each little interval of depth.

\item At first we assume that gravity is
  so weak that we can ignore it and assume that the pearls stop completely. Then in the second calculation we suppose that the pearls continue to fall directly downwards under gravity with a constant very low velocity. After the set in of this ``terminal velocity'' the pearls contribute the same amount of radiation in all depth layers below the point where the
  ``terminal velocity'' has set in. This means that we consider that
   there are two different stages in the motion of a dark matter particle:

  \begin{itemize}
  \item A fast stage from when the dark matter particle enters the
    Earth through its surface until the particle has slowed down and
    we assume it moves so fast that it has no time to emit significant
    amounts of radiation. So during this stage we can neglect the radiation
    and the particle is effectively invisible.
  \item In the next stage we first consider the case where
   the dark matter particle
    stops and stays sitting in the Earth until its amount of excitation has
    burned out (after a long time, say a week, but taken at least as an
    average as a fixed constant). This case is illustrated by the {\em black} rectangle in Figure \ref{dist}.
    
  \end{itemize}
\begin{figure}
	\includegraphics[scale=0.7]{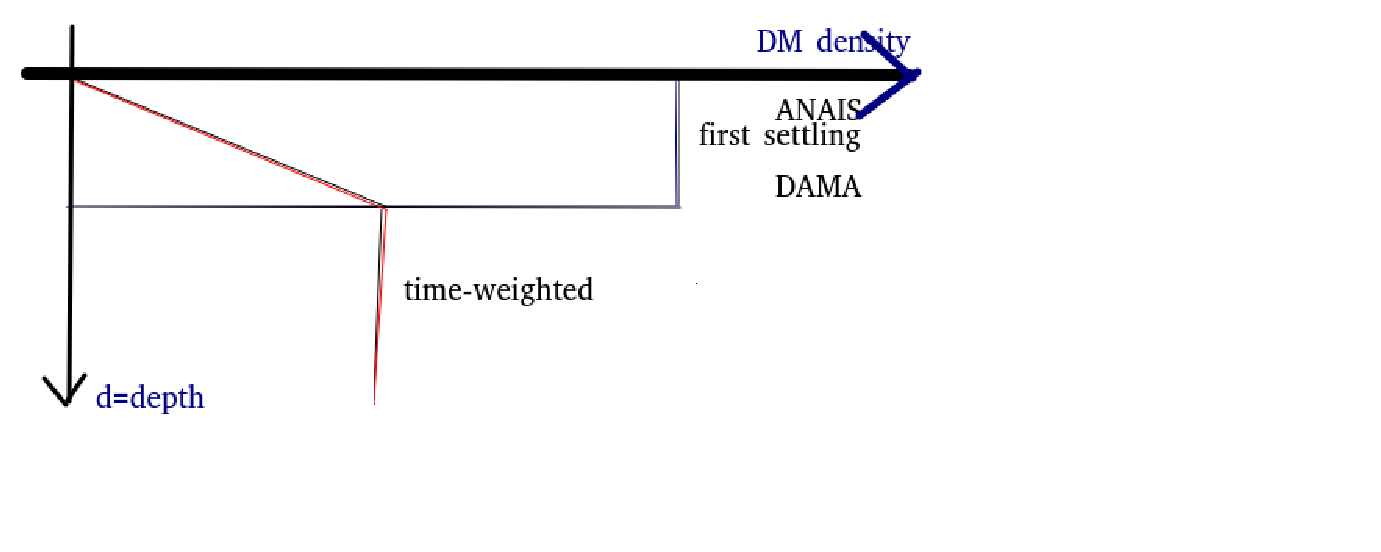}
	\caption{\label{dist} 
		We consider a model in which the incoming direction of the dark matter pearls has a constant distribution over the sky. At first we assume they stop completely after passing through a "stopping length" of earth and only then radiate significantly.
		Then we get a flat distribution
		of radiation delivery as a function of the depth down to a
		deepest depth equal to the "stopping length", beyond which there is no more probability for finding any
		radiation. This corresponds to the black rectangle on the figure.
		However if the pearls do not stop completely, but begin to fall with a constant velocity
		determined by a balance between the friction and the gravitational
		attraction by the Earth after reaching the "stopping length", then
		the distribution of the radiation delivered is given by the red
		curve as a function of the depth.}
\end{figure}

\item Further, as a simplifying assumption, we assume that all the particles
  have the same stopping length (in the earth) and have the same incoming velocity.
  So all the dark matter particles hitting the Earth's surface at a certain point end up and stop (or in the second case considered below go into the next stage with a terminal velocity) at a point
  lying on a half sphere around the point of impact with the Earth, having a radius equal to the stopping length.
\item Again for simplicity, we assume that the distribution of the
  directions from which the dark matter particles come is evenly distributed
  over the sky. But it is of course in our model with a stopping length
  much shorter than the radius of the Earth, only the half sphere where the
  particles going downwards can come from that gets populated. Particles seeking to
  come upwards are stopped on the other side of the Earth.
\item We also assume that the radiation (of electrons or X-rays
  of energy 3.5 keV) emitted during the first stage when the pearls pass quickly through the medium (essentially the earth, or the
  experimental apparatus scintillator) is negligible compared to the radiation emitted at the stopping point or, in the second case, 
  when the dark matter sinks (before it runs out of
  excitation energy). 
  \end{itemize}

Let us now illustrate the rather trivial calculation of the
radiation intensity observed:

First notice that if the dark matter particle comes in with a direction making an angle $\theta$ with the downward vertical, then the depth to which it has reached when the particle has stopped is
\begin{eqnarray}
  \hbox{stopping depth } &=& ``stopping\; length''*\cos(\theta)
\end{eqnarray}

Now the area on the half sphere - which by the assumption of an even
distribution on the sky is proportional to the fraction of the dark matter
particles ending up there - corresponding to the depth $z$ being in the
infinitesimal interval $dz$ is given as
\begin{eqnarray}
  dz&=& ``stopping \; length'' *d\cos\theta\\ &=&
  (-) ``stopping \; length'' *\sin\theta*d\theta
  \end{eqnarray}
 while	
 \begin{eqnarray}
  d``area''&=& 2\pi*(``stopping \; length'')^2 *d\theta *\sin{\theta}.
  \end{eqnarray}
  so that 
  \begin{eqnarray} 
   dz &\propto& d``area'' \propto d`` probability''\,
  \hbox{up to $z$ = $`` stopping \; depth''$}.
  \end{eqnarray}
This means that the radiation is present only down to the depth
just equal to the stopping length, and in the interval with
lower depth than that the radiation rate is quite constant. This is the black rectangular distribution
given in Figure \ref{dist}, which is
namely the case in which the dark matter particles sit still on the surface
of the half sphere and radiate out.

In the second case the dark matter particles almost stop and then start moving straight downwards with a constant terminal velocity, which is so slow that they contribute significant amounts of radiation at lower depths.
Hence one gets the radiation rate for a depth $z$ to be given by the integral of the rectangular distribution over all the higher depths:

\begin{eqnarray}
  ``Rate''( z) &\propto& \int_{0}^{z} ``rate''_{rectangular}(z') dz',
  \end{eqnarray}
It is easy to see that the resulting
radiation rate $``Rate''(z)$ has a non-zero slope in the
uppermost stopping length, while it becomes constant under this domain.
\begin{figure}
	\includegraphics[scale=0.7]{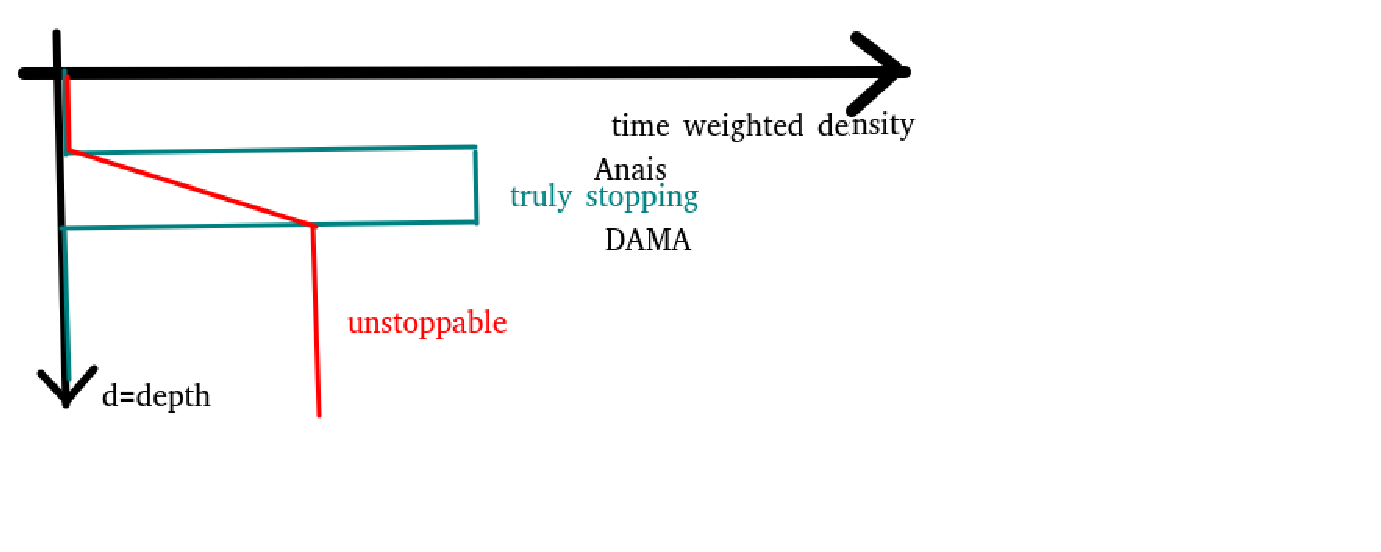}
	\caption{\label{wg} Here we take into account that the Earth's gravity
		acts on the dark matter pearl even while it is in the fast stage of its motion. Then even a pearl
		coming in almost horizontally will sink a bit before it stops or falls with its terminal velocity
		So in fact the black rectangle from
		figure \ref{dist} will essentially be lowered by a constant amount
		in depth. Also the distribution for the delivery of the radiation in
		the case of the pearl having a terminal velocity gets a corresponding lowering
		of the red curve from figure \ref{dist}.}
\end{figure}


  Next we want to repeat the above two calculations including the effect of  gravitation under the fast stage, which we have neglected so far. We perform this
  improvement very crudely assuming:

  \begin{itemize}
  \item We can ignore the variation as a function of the angle $\theta$
    of the time the stopping in the fast stage takes. I.e. whatever the angle
    and thereby the depth in which the particle essentially stops it takes
    the same time, counted from the passage through the Earth's surface.
  \item Much like in an Einstein elevator the particle falls compared to the
    motion ignoring gravity by the amount that a free fall in the same
    time would have caused an object to fall.
  \end{itemize}

  With these assumptions the effect of introducing gravity on the curves
  of Figure \ref{dist} is simply to move them downward by the free fall
  distance. We here mean the free fall distance of a body falling for as long as the
  stopping time. The uppermost range of depth corresponding to this falling
  distance, becomes completely free of radiation in our approximations.
  So an experiment so close to the surface should see no dark matter at all.

  These moved down curves are drawn in Figure \ref{wg}.

  If we could arrange say that ANAIS should lie so close to the surface
  as the falling distance during the stopping time then, under the rather simplified assumptions just presented, we could explain the lack of
   observation of dark matter.






 
\section{Size}

  The dragging force stopping the dark matter pearl is
  \begin{eqnarray}
    F_D &=&C_D \frac{A v^2 \rho_{material \; outside \; pearl}}{2}
  \end{eqnarray}
  where $v$  is its velocity and
  $ \rho_{material \; outside \; pearl}$ the density of the fluid or material
  through which the pearl falls. $C_D$ is the drag coefficient and is
  of order unity at high speed. $A\approx \sigma$ is the area shown to the
  motion.

  The equation of motion (Newton's second law) becomes:
  \begin{eqnarray}
    M\dot{v}&=& -F_D =-\frac{C_D \sigma v^2\rho_{material \; outside \; pearl}}{2}
    \end{eqnarray}
which can be rewritten and integrated to give the stopping length $L_{stopping}$ as follows:
  \begin{eqnarray}
    \frac{\dot{v}}{v^2}&=& -\frac{\sigma}{M} *\frac{C_D}{2}
    \rho_{material \; outside \; pearl}\\
    -\frac{1}{v}&=& -t*\frac{\sigma}{M} *\frac{C_D}{2}
    \rho_{material \; outside \; pearl}+const.\\
    v&=& \frac{1}{t*\frac{\sigma}{M} *\frac{C_D}{2}
      \rho_{material \; outside \; pearl}-const.}\\
    L_{stopping}=\int vdt &=&
    \int_{v=300km/s}^{v \; small}\frac{dt}{t*\frac{\sigma}{M}
      *\frac{C_D}{2}\rho_{material \; outside \; pearl}-...}\\
    &\approx&- (\frac{\sigma}{M} *\frac{C_D}{2}
    \rho_{material \; outside \; pearl})^{-1}\ln(\frac{``small''}{300km})\nonumber
    \end{eqnarray}


   
  Rewriting the estimate of the stopping length we have
  \begin{eqnarray}
  L_{stopping}&\approx& \frac{1}{\frac{\sigma}{M} *\frac{C_D}{2}
  \rho_{material \; outside \; pearl}\ln(\frac{300km/s}{``small''})}\nonumber\\
  L_{stopping}\rho_{material \; outside \; pearl}\sigma &\approx&
    M\frac{2}{C_D}\ln(\frac{300km/s}{``small''}) \label{Lstop}
\end{eqnarray}
 Now  $L_{stopping}\sigma*\rho_{material \; outside \; pearl}$ is the amount of material pressed away by the passage of the pearl through the earth.
So noticing that $\frac{2}{C_D}*\ln \frac{300 km/s}{``small''}$ is just
of order unity, say $30$, we see that the mass of the material pressed away by the 
the pearl  is only an order of unity times bigger than the
mass $M$ of the pearl itself.


If the pearl essentially stops at the DAMA depth with 4200m w.e. then we have
  \begin{eqnarray}
    4200 m *1000kg/m^3 &=&\frac{M}{\sigma} \frac{2}{C_D}*
    \ln\frac{300km/s}{``small''}\\
    \hbox{giving }\frac{\sigma}{M} &=& \frac{1}{4.2*10^6kg/m^2 *30}\\
    &=& 10^{-7}m^2/kg.
\end{eqnarray}
This should be compared with the inverse darkness obtained by Correa for velocities around 300 km/s from her analysis of dwarf galaxies \cite{CAC}
\begin{eqnarray}
    \frac{\sigma}{M}&=& 1cm^2/g =0.1m^2/kg.
  \end{eqnarray}
  This means we need that so much dust washed off  
   the pearls coming into the Earth 
  that their cross section is diminished by a factor 1 million.
  So in area, they should have lost a factor $10^6$
  and in linear scale a factor $10^3$.
  
\section{Physics of hoped for phase transition}

  By fitting to our dark matter model we found the order of magnitude
  of the tension $S$ of the domain wall between the two phases should lie in the range
  \begin{eqnarray}
    S&=& (\hbox{few}\ MeV)^3 \ \hbox{to say } (30\, MeV)^3
  \end{eqnarray}

  This indicates that the physics involved in making this two vacua,
  if it is right, should be in an energy range which is {\em not at all new}.

  The number $(30MeV)^3$ is what letting the domain walls replace the
  cosmological constant (the dark energy) would require \cite{Corfu2022A}.

  We obtained the relationship between the mass $M$ and surface tension $S$ for our pearls mainly by adjusting the density of the strongly compressed material
  of ordinary matter inside the new type of vacuum to have a HOMOLUMO gap
  suitable for emitting a 3.5 keV X-ray line. We further assumed that the pearls are not far from collapsing and spitting out the nuclei inside, in spite of a
  potential keeping them in of $\Delta V \approx 3MeV$. This relationship is illustrated here by giving examples of the mass M for a few values of 
 the cubic root of the surface tension $S$.
  
   \begin{eqnarray}
 	S^{1/3}= 1\,  GeV&\Rightarrow& M = 24\, ton\\
 	\hbox{while \quad }S^{1/3}= 100\, MeV &\Rightarrow& M=24\, mg\\
 	\hbox{and \quad }S^{1/3}=10\, MeV &\Rightarrow&
        M=2.4 *10^{-14}\,kg=1.4 *10^{13}\, GeV.
   \end{eqnarray}

   If we take the tension to be say $(10MeV)^3$, then we should look for physics at
   this scale to find out what could be used to cause the phase transition making
   the two vacua.


\section{Conclusion}

We have briefly reviewed our model for the dark matter
being pieces of a new vacuum phase - which should though be understandable
in terms of Standard Model physics, namely the Nambu JonaLasinio chiral
symmetry breaking and QCD - filled under high pressure with ordinary matter.

But mainly we have searched for a way to resolve the seemingly very hard
mystery that, while the DAMA-LIBRA experiment has collected
more and more evidence for dark matter other experiments have not yet seen any. In particular the ANAIS and COSINE-100 experiments are very similar in that they also use  NaI(Tl) scintillator but do not confirm the seasonal variation in their data which DAMA-LIBRA use to single out the dark matter. AMAIS and COSINE-100 claim a 3 $\sigma$ disagreement with DAMA-LIBRA but COSINE-100 has less data.
Even more, all the xenon scintillator based experiments
having searched for dark matter have seen nothing.

The solution we have proposed to this disagreement is that dark matter is macroscopic and interacts more strongly than assumed in other models such as WIMP theories, but still sufficiently dark to fit 
observations. WIMPS are not significantly stopped by the earth shielding. However our dark matter pearls are stopped in a depth of order of that of the DAMA-LIBRA experiment 1400m, after passing through the Earth's surface with a high galactic velocity of the order of 200 km/s. In our model the
dark matter can be excited and then emit electrons or photons with
the preferred energy of 3.5 keV (corresponding to the controversial X-ray
line associated with dark matter). So we can claim that the main effect seen by
DAMA-LIBRA is this 3.5 keV radiation emitted by stopped
or very much slowed down dark matter particles. But  the dark matter
particles pass through the experiments like ANAIS and COSINE-100 so fast that there is not sufficient time for them to give an observable signal. If really we take the dark matter
to only be effectively observable when stopped and having time to emit
its excitation energy, then the {\em fluid} xenon used by the majority of
the underground experiments looking for dark matter cannot keep the dark matter
stationary and thus, under such assumptions, have no chance to ``see'' the dark
matter. At least if the Earth's gravitation is sufficient to drive the dark matter
particles through the liquid xenon the liquid xenon experiments cannot
observe them.

Very crucial for our speculation is that the signal DAMA-LIBRA sees is indeed 3.5 keV radiation emitted by stopped dark matter particles. It is a {\em remarkable fact} that the DAMA-LIBRA spectrum of the seasonally varying component is fitted
by an essentially Gaussian distribution in energy with an average
energy {\em $3.15\, keV$ close to the $3.5\, keV$ line}.

We remark that in our model we can even obtain a negative seasonal variation (i.e. with more events in winter than in summer). Interestingly COSINE-100 generated such an effect, which they consider as spurious, by using the DAMA-LIBRA background subtraction procedure on their data.

But if now dark matter is indeed, as we suggest, stopped
in a depth under the Earth's surface of the order of the depth of DAMA, 1400 m,
then one should - of course - seek to {\em dig dark matter out at this depth.}
In water equivalent depth it would accidentally be very close to the
bottom of the oceans, 5 km of water. The dark matter should be easily distinguished
by having an abnormally high specific weight. 

So finally we believe that the DAMA-LIBRA results, contrary to other experiments, 
point to the need for a macroscopic model of dark matter and that the mean free path
in earth for galactic velocity dark matter must be of the order of 1400 m.

\section*{Acknowledgement}
One of us H.B.N. acknowledges emeritus status at the Niels Bohr Institute, and
 support to the tour to Bled.

\end{document}